\documentclass{aastex}          %% The manuscript based on AASTeX v5.x
\usepackage{spr-astr-addons-fixed2}    %% mimicing ASTR journal style
\begin{document}
%% Article title
%
%\title{Status report on the EXOTIME program}
\title{EXOTIME: searching for planets around pulsating subdwarf B stars}

%% Running heads
\shorttitle{EXOTIME: searching for planets around pulsating subdwarf B stars}
\shortauthors{Schuh, Silvotti, Lutz et al.}

%% Author and Affilations
\author{Sonja Schuh\altaffilmark{1,2}} 
%\affil{blubbs?}
\email{schuh@astro.physik.uni-goettingen.de} %% non-output
\and 
\author{Roberto Silvotti\altaffilmark{3,4}}
%\affil{blubbs?}
\and
\author{Ronny Lutz\altaffilmark{2,5}} 
\and
\author{Bj\"orn Loeptien\altaffilmark{2}} 
\and
\author{Elizabeth M.\ Green\altaffilmark{6}}
\and
\author{Roy H.\ {\O}stensen\altaffilmark{7}}
\and
\author{Silvio Leccia\altaffilmark{4}}
\and
\author{Seung-Lee Kim\altaffilmark{8}}
\and
\author{Gilles Fontaine\altaffilmark{9}}
\and
\author{St{\'e}phane Charpinet\altaffilmark{10}}
\and
\author{Myriam Franc{\oe}ur\altaffilmark{9}}
\and
\author{Suzanna Randall\altaffilmark{11}}
\and
\author{Cristina Rodr{\'\i}guez-L{\'o}pez\altaffilmark{10,12}}
\and
\author{Valerie van Grootel\altaffilmark{10}}
\and
\author{Andrew P. Odell \altaffilmark{13}}
\and
\author{Margit Papar{\'o}\altaffilmark{14}}
\and
\author{Zs\'ofia Bogn{\'a}r\altaffilmark{14}}
\and
\author{P\'eter P{\'a}pics\altaffilmark{7,15}}
\and
\author{Thorsten Nagel\altaffilmark{1}}
\and
\author{Benjamin Beeck\altaffilmark{2}}
\and
\author{Markus Hundertmark\altaffilmark{2}}
\and
\author{Thorsten Stahn\altaffilmark{5}}
\and
\author{Stefan Dreizler\altaffilmark{2}}
\and
\author{Frederic V. Hessman\altaffilmark{2}}
\and
\author{Massimo Dall'Ora\altaffilmark{4}}
\and
\author{Dario Mancini\altaffilmark{4}}
\and
\author{Fausto Cortecchia\altaffilmark{4}}
\and
\author{Serena Benatti\altaffilmark{16}}
\and
\author{Riccardo Claudi\altaffilmark{17}}
\and
\author{Rimvydas Janulis\altaffilmark{18}}

%% Alternate Affilations
%\altaffiltext{}{} %dummy
\altaffiltext{}{} %dummy
\altaffiltext{}{} %dummy
\altaffiltext{1}{Institut f\"ur Astronomie und Astrophysik, Kepler Center for Astro and Particle Physics, Eberhard-Karls-Universit\"at, 
  Sand 1, 72076 T\"ubingen, Germany}
\altaffiltext{2}{Georg-August-Universit\"at G\"ottingen, Institut f\"ur Astrophysik, 
              Friedrich-Hund-Platz~1, 37077 G\"ottingen, Germany}
\altaffiltext{3}{Istituto Nazionale di Astrofisica (INAF), Osservatorio Astronomico di Torino, Strada Osservatorio 20, 
  10025, Pino Torinese, Italy}
\altaffiltext{4}{Istituto Nazionale di Astrofisica (INAF), Osservatorio Astronomico di Capodimonte, via Moiariello 16,
              80131 Napoli, Italy}
\altaffiltext{5}{Max-Planck-Institut f\"ur Sonnensystemforschung, Max-Planck-Straße 2, 37191 Katlenburg-Lindau, Germany}
\altaffiltext{6}{Steward Observatory, University of Arizona, 933 North Cherry Avenue, Tucson, AZ 85721, USA}
\altaffiltext{7}{Instituut voor Sterrenkunde, K. U. Leuven, Celestijnenlaan 200D, 3001 Leuven, Belgium}
\altaffiltext{8}{Korea Astronomy and Space Science Institute, Daejeon 305-348, Korea}
\altaffiltext{9}{D{\'e}partement de Physique, Universit{\'e} de Montr{\'e}al, C.P. 6128 Succ. Centre-Ville, Montr{\'e}al, H3C 3J7, Canada}
\altaffiltext{10}{Laboratoire d'Astrophysique de Toulouse-Tarbes, Universit{\'e} de Toulouse, CNRS, 14 Avenue {\'E}douard Belin, 31400 Toulouse, France}
\altaffiltext{11}{ESO, Karl-Schwarzschild-Str.\ 2, 85748 Garching bei M\"unchen, Germany}
\altaffiltext{12}{Departamento de F{\'\i}sica Aplicada, Universidade de Vigo, Vigo 36310, Spain}
%\altaffiltext{13}{Departmento de F{\'\i}sica Estelar, Instituto de Astrof{\'\i}sica de Andaluc{\'\i}a-CSIC, Granada 18008, Spain}
\altaffiltext{13}{Northern Arizona University, Flagstaff, AZ 86011, USA}
\altaffiltext{14}{Konkoly Observatory of the Hungarian Academy of Sciences, P.O.\ Box 67., H-1525 Budapest XII, Hungary}
\altaffiltext{15}{E\"otv\"os Lor{\'a}nd University Faculty of Science, P{\'a}zm{\'a}ny P{\'e}ter s{\'e}t{\'a}ny 1/A, H-1117 Budapest, Hungary}
\altaffiltext{16}{CISAS, Universit{\`a} degli studi di Padova, via Venezia 15, 35131 Padova, Italy}
\altaffiltext{17}{Istituto Nazionale di Astrofisica (INAF), Osservatorio Astronomico di Padova, vicolo Osservatorio 5, 35122 Padova, Italy}
\altaffiltext{18}{Institute of Theoretical Physics and Astronomy, Vilnius University, 12 A. Gostauto Street, 01108 Vilnius, Lithuania}
%\vspace{-15mm}

%% Abstract
\begin{abstract}
In 2007, a companion with planetary mass was found around the
pulsating subdwarf B star V391~Pegasi with the timing method,
indicating that a previously undiscovered population of substellar
companions to apparently single subdwarf B stars might
exist. Following this serendipitous discovery, the
\mbox{EXOTIME}\footnote{\texttt{http://www.na.astro.it/$\sim$silvotti/exotime/}}
monitoring program has been set up to follow the pulsations of a
number of selected rapidly pulsating subdwarf B stars on time-scales
of several years with two immediate observational goals:

1) determine $\dot{P}$ of the pulsational periods $P$

2) search for signatures of substellar companions in O$-$C residuals due 
to periodic light travel time variations, which would be tracking the central star's 
companion-induced wobble around the center of mass.

These sets of data should therefore at the same time: on the one hand
be useful to provide extra constraints for classical
asteroseismological exercises from the $\dot{P}$ (comparison with
''local'' evolutionary models), and on the other hand allow to
investigate the preceding evolution of a target in terms of possible
''binary'' evolution by extending the otherwise unsuccessful search
for companions to potentially very low masses.

While timing pulsations may be an observationally expensive method to
search for companions, it samples a different range of orbital
parameters, inaccessible through orbital photometric effects or the
radial velocity method: the latter favours massive close-in companions,
whereas the timing method becomes increasingly more sensitive towards 
wider separations. 

In this paper we report on the status of the on-going observations and
coherence analysis for two of the currently five targets,
revealing very well-behaved pulsational characteristics in
\mbox{HS\,0444+0458}, while showing \mbox{HS\,0702+6043} to be more
complex than previously thought.
\end{abstract}

%% Keywords

\subsection*{}

\small%
Note: \textit{This paper contains the content of three
contributions originally presented %during the conference
as}
\begin{itemize}
\item {Status report on the EXOTIME program} 
  % \textit{(Talk, Schuh)}
\item {EXOTIME photometric monitoring of HS\,0444+0458} 
  % \textit{(Poster, Schuh et al.)}
\item {EXOTIME photometric monitoring of HS\,0702+6043}.
  % \textit{(Poster, Lutz et al.)}
\end{itemize}
\normalsize
\vfill

%%  Please use labels (\label, \ref) for section, figure, table, 
%%  equation  reference. Use \cite for bibliography references.
%
%
\section{Introduction}\label{s:intro}
While $\dot{P}$ measurements exist for a number of pulsating white
dwarfs
\citep[e.g.][]{1991ApJ...378L..45K,2000ApJ...534L.185K,2005ApJ...634.1311K,1999ApJ...522..973C,2008A&A...489.1225C,2008ApJ...676..573M},
similar measurements have only been published for one pulsating
subdwarf B star so far, \object{V391~Pegasi} (hereafter
\object{HS\,2201+2610}).

\object{HS\,2201+2610} was first discovered to be a rapidly pulsating
subdwarf B star by
\citet{2001A&A...368..175O}. \citet{2002A&A...389..180S,2007Natur.449..189S}
were able to derive $\dot{P}$ values for the two strongest pulsation
modes, and found an additional pattern in the O$-$C diagrams that
revealed the presence of a giant planet in a 3.2 year orbit.
Additional slow pulsations were subsequently reported by
\citet{2009A&A...496..469L}.

From theoretical considerations, both the rapid p-mode as well as
the slow g-mode oscillations should be very stable in the pulsating
subdwarf B stars (stable in phase, i.e.\ coherent over time scales of
many years, as well as stable in amplitude) due to the identical
underlying driving: a $\kappa$ mechanism acting on the $Z$
bump. Time dependency of phases and amplitudes can however be 
introduced through nonlinear effects; the resulting mode coupling
then starts to invalidate the initial simple assumption. It is 
therefore crucial to identify target stars that show a stable 
behaviour.

Under the right conditions,
this provides the possibility to obtain meaningful results from
long-term photometric monitoring of rapidly pulsating sdB stars: It
allows to search for slow changes in pulsation periods due to
evolutionary effects, testing the time scales in sdB evolutionary
models, and may be used as an additional constraint for sdB
asteroseismological models. It also allows to use the star as a clock
that provides a regular timing signal, which can be exploited to
search for small periodic deviations from the mean pulsation maxima
arrival times. Such periodic deviations can be caused by a varying
light travel time and may hence indicate a wobble of the pulsating
subdwarf B star's location due to one or more low-mass companions, as
is the case for \object{HS\,2201+2610}. This timing method is
sensitive to planetary masses and large period ranges not easily
accessible with other methods.

Given the observational efforts required, the detection rate of
low-mass companions in wide orbits is surprisingly high so
far. Besides timing of pulsations \citep{2007Natur.449..189S}, timing
of the sharp eclipses in HW~Vir type systems has been used to detect
tertiary bodies \citep{2009AJ....137.3181L,2009ApJ...695L.163Q}.
Furthermore, \citet{2009ApJ...702L..96G} have even successfully
employed the radial velocity method to detect a close substellar
companion to \object{HD 149382}.

\begin{table*}[t]
%\begin{center}
\small
\caption{Overview of EXOTIME targets} %% no full stop at the end of caption
\label{tbl:targets}
    \begin{tabular}{lllll}
      \tableline
      target             & coordinates (equinox 2000.) & $m_{\textrm{B}}$ & status & \\[1mm]
      \tableline
      {HS\,2201+2610}    & 22:04:12.0 \quad +26:25:07 &14.3& {collecting data} &
                                     {planet candidate published}, 
                                     {$\sin{i}$~unknown}  \\[1mm]
      {HS\,0702+6043}    & 07:07:09.8 \quad +60:38:50 &14.7& {collecting data}&{see \ref{ss:hs0444} and \ref{s:longterm}} \\[1mm]
      {HS\,0444+0458}    & 04:47:18.6 \quad +05:03:35 &15.2 & {collecting data}&{see \ref{ss:hs0702}} \\[1mm]
      EC\,09582$-$1137~  & 10:00:41.8 \quad $-$11:51:35 &15.0& {collecting data}& \\[1mm]
      PG\,1325+101       & 13:27:48.6 \quad +09:54:52 &13.8& {collecting data}& \\[1mm]
      \tableline
    \end{tabular}
%\end{center}
\end{table*}

As it has been suggested that the influence of substellar companions
or large planetary companions may be decisive for the evolution of
subdwarf B stars (see Discussion in section~\ref{s:discussion}), the
EXOTIME program has been set up to find more such objects.  We discuss
the target selection in section~\ref{s:data}.  However, as amplitude
variations have repeatedly been reported in sdB pulsators
\citep[e.g.][]{2010ApSS.K}, it is crucial to carefully check the
stability of the targets first of all. This contribution mainly
discusses the current status of these investigations for
two of our objects, \object{HS\,0444+0458} and \object{HS\,0702+6043}
(section~\ref{s:coherence}).

\object{HS\,0444+0458} was first discovered to pulsate by
\citet{2001A&A...378..466O}, and has been further characterised by
\citet{2007MNRAS.378.1049R}.
Rapid oscillations were discovered in \object{HS\,0702+6043} by
\citet{2002A&A...386..249D}, simultaneous slow oscillations were
reported by \citet{2006A&A...445L..31S}. The on-going \mbox{EXOTIME}
observations for \object{HS\,0702+6043} have also previously been
summarised by \citet{2008CoAst.157..185L,2009CoAst.159...94L}.

%These two and two further objects were initially selected as targets for
%the EXOTIME program according to the considerations in the next section.

\section{Observational Data}\label{s:data}
\subsection{Target selection}\label{ss:targets}

Similar to the criteria applied when first selecting
\object{HS\,2201+2610} for long-term monitoring, we considered two sets
of characteristics to choose further targets. The first set concerns
the pulsations: frequencies, amplitudes and phases are obviously
required to be stable over long periods of time. Although there are
early indicators if these criteria are not fulfilled, it should be
clear these properties can only be truly verified, in particular for
the phases, from long-term monitoring data in hindsight (see Section
\ref{s:coherence}). 

Using target properties partly taken from
\citet{2007MNRAS.378.1049R}, we selected candidates with few but
preferably more than one or two frequencies, in order to be able to
fully resolve the frequency spectrum during short individual runs
while also retaining the possibility to obtain an independent O$-$C
measurement from each individual frequency. For the amplitudes we
leaned towards medium values, avoiding, where possible, very small
amplitudes which would have S/N issues, as well as very large ones
where one might have to start worrying about non-linearity effects.

We also used \citeauthor{2007MNRAS.378.1049R}'s listing to consider
the amplitude stability criterion given by\linebreak
\citet{2001ApJ...562L.141C}, which for oscillations similar to
solar-like ones identifies the \emph{stochastically excited} pulsations.
While this quantitative distinction for the underlying type of driving
has, not surprisingly at all, been previously shown inadequate for the 
\emph{opacity-driven} subdwarf B pulsators, we still found it helpful to 
restrict our list of objects to those with
$\sigma(A) / \langle A\rangle < 0.5 $.
%$\displaystyle \frac{\sigma(A)}{\langle A\rangle} < 0.5 $,
We took this as a purely observational value that overall should favour 
the more stable pulsators.

The second set of selection criteria was concerned with more practical
questions, looking for the brighter ones of the available candidates
and including preferentially northern hemisphere targets. Compromising
between all of these considerations, we selected the objects in
Table~\ref{tbl:targets} as the initial list of targets to be observed.

\subsection{Observations}\label{ss:observations}

Due to not only the different magnitudes and pulsation amplitudes as
well as the spatial distribution of the targets, but also due to the
long-term nature of the project and the changing availability of
telescopes, the observations in the data bases for our target objects
come from a variety of sites, instruments, and observers. The bulk of
observations for \object{HS\,2201+2610} have been published with
extensive observing logs by
\citet{2002A&A...389..180S,2007Natur.449..189S}. Further data is still
being collected and analysed in order to confirm the future evolution
in the O$-$C diagram as predicted from the known orbital motion, as well
as to search for possible further companions at different orbital
periods.

The two other targets with the best coverage so far (after
\object{HS\,2201+2610}) are \object{HS\,0702+6043} and\linebreak
\object{HS\,0444+0458}. For these two targets, the full observing logs
up to date are provided in Table~\ref{tbl:hs0444}
(\object{HS\,0444+0458}) and Table~\ref{tbl:hs0702a}
(\object{HS\,0702+6043}), respectively.
%include $\sigma(A)$ in tables
The observations available for these two targets come from telescopes
ranging from 0.5\,m to 3.6\,m in aperture, and were all obtained with
CCD detectors, almost all of the time through B filters, and 
with exposure times between 5\,s and 40\,s.

%%%%%%%%%%%%%%%%%%%%%%%%%%%%%%%%%%%%%%%%%%%%%%%%%%%%%%%%%%%%%%%%%%%%%%%%%%%%%%%%%%%%%%%%%%
%% Figures
%
\begin{figure}[tb]
\includegraphics[width=\columnwidth,angle=0]{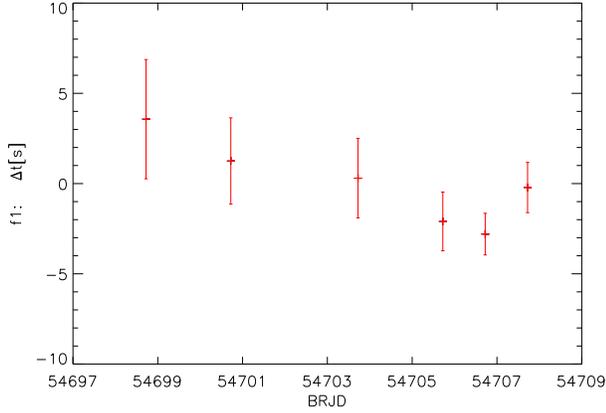}
\caption{TNG data on {HS\,0444+0458}: O$-$C diagram for f1 \qquad ~} %% no full stop at the end of caption
\label{fig:hs0444tngocf1}
\end{figure}

\begin{figure}[tb]
\includegraphics[width=\columnwidth,angle=0]{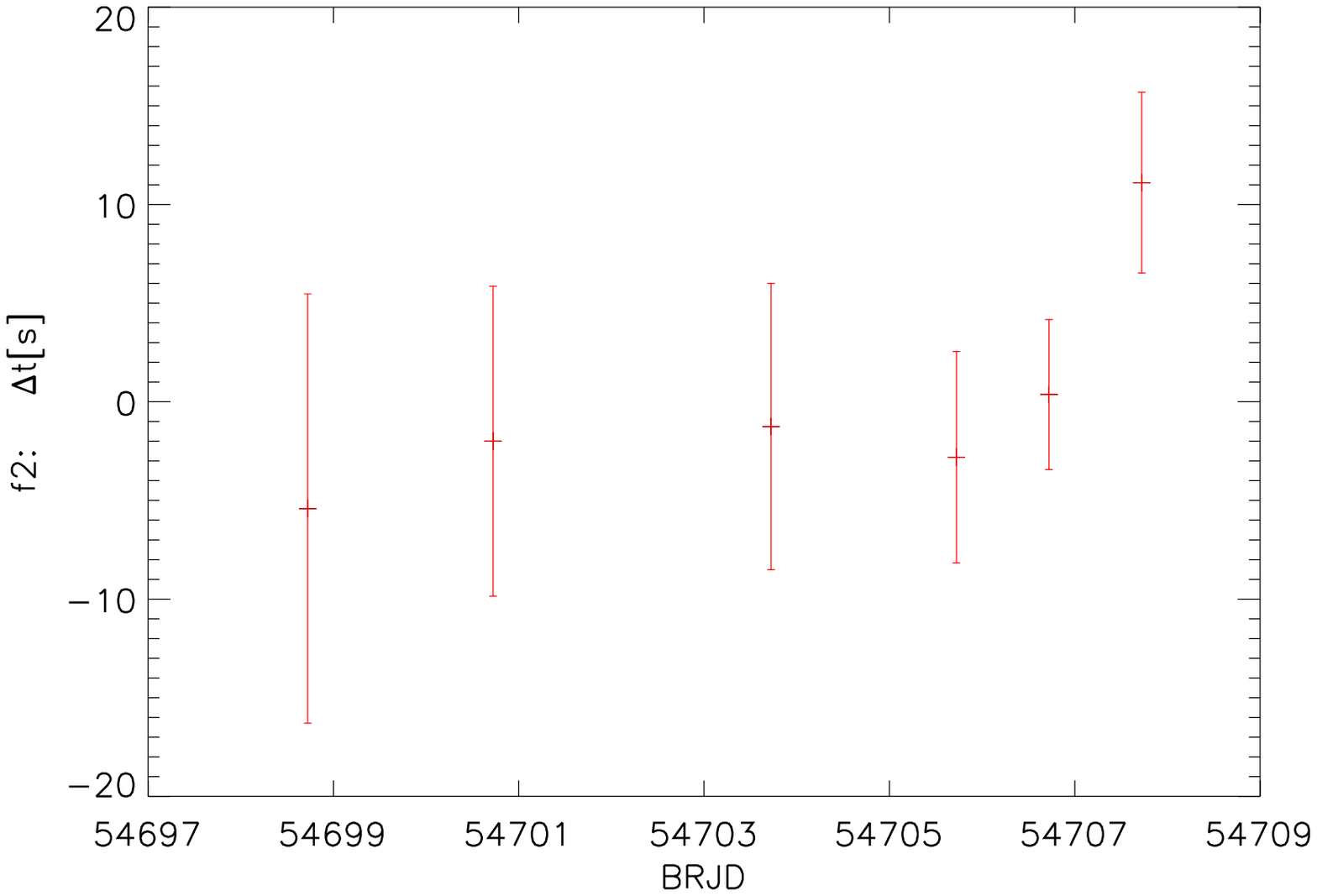}
\caption{TNG data on {HS\,0444+0458}: O$-$C diagram for f2 \qquad ~} %% no full stop at the end of caption
\label{fig:hs0444tngocf2}
\end{figure}

\begin{figure}[tb]
\includegraphics[width=\columnwidth,angle=0]{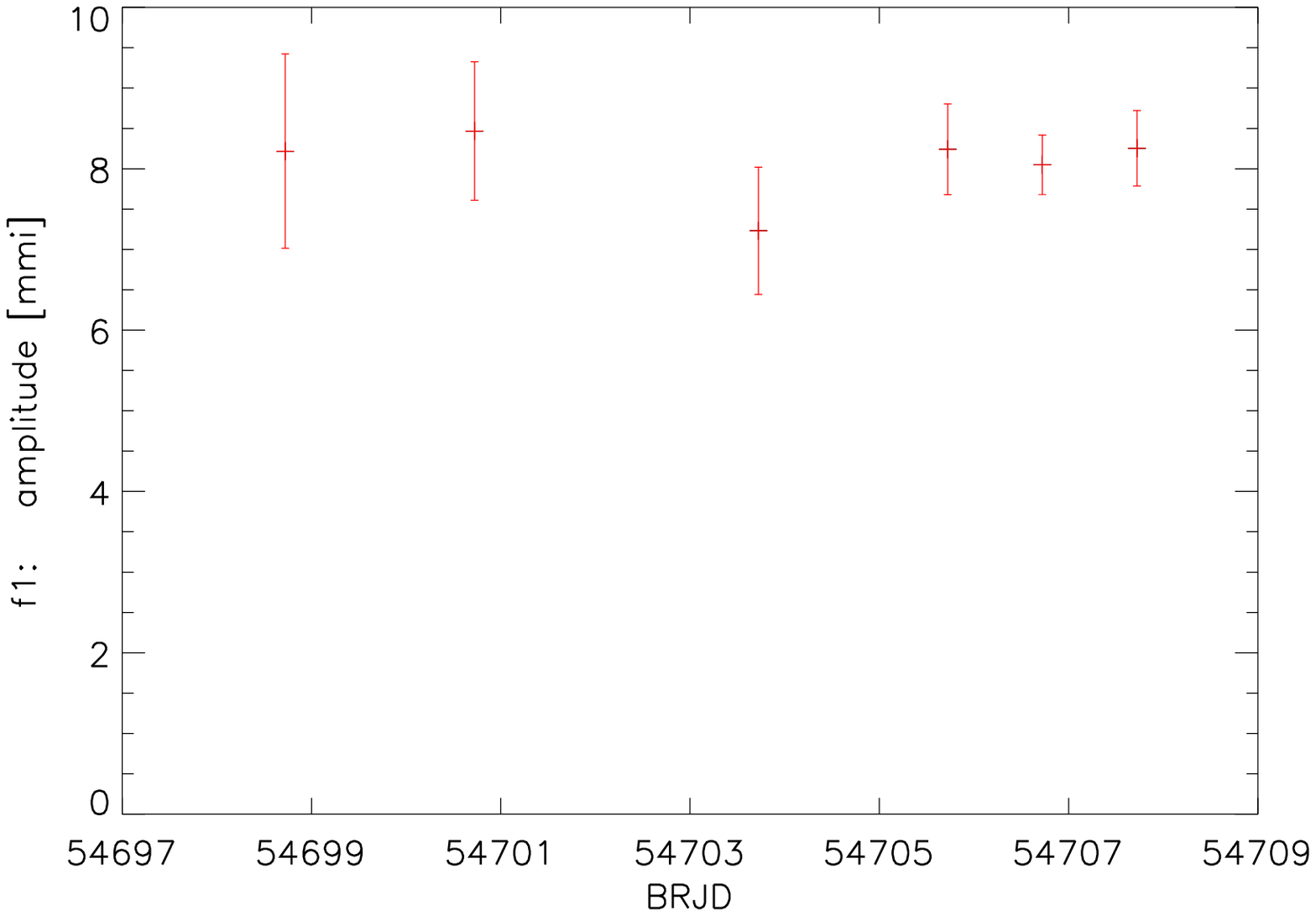}
\caption{TNG data on {HS\,0444+0458}: amplitude variation for f1} %% no full stop at the end of caption
\label{fig:hs0444tngampf1}
\end{figure}

\begin{figure}[tb]
\includegraphics[width=\columnwidth,angle=0]{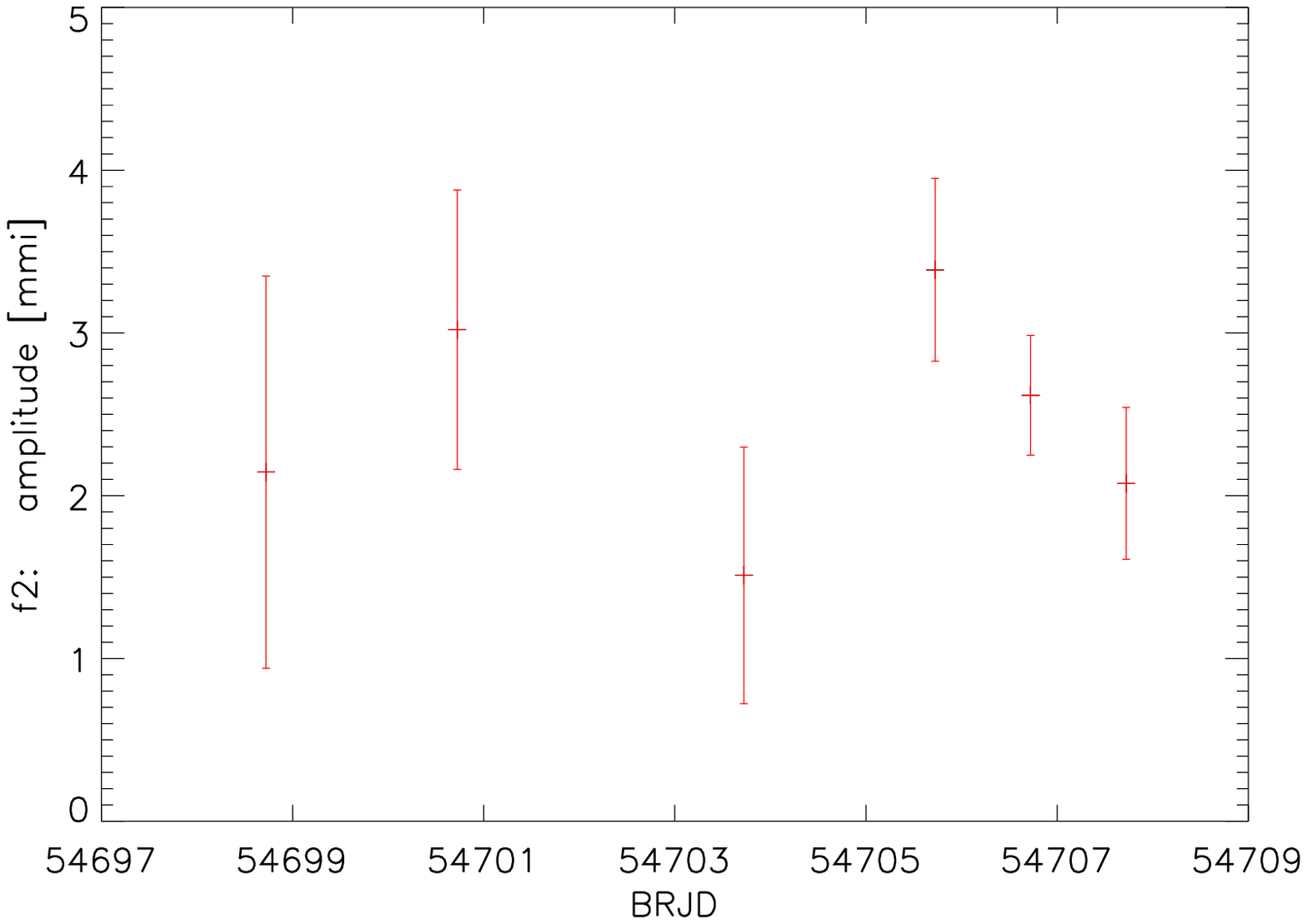}
\caption{TNG data on {HS\,0444+0458}: amplitude variation for f2} %% no full stop at the end of caption
\label{fig:hs0444tngampf2}
\end{figure}

%%%%%%%%%%%%%%%%%%%%%%%%%%%%%%%%%%%%%%%%%%%%%%%%%%%%%%%%%%%%%%%%%%%%%%%%%%%%%%%%%%%%%%%%%%

\begin{table}[b]
\small
\caption{The two main frequencies in {HS\,0444+0458} as obtained from the full 2008-2009 data set} %% no full stop at the end of caption
\label{tbl:hs0444frequencies}
    \begin{tabular}{@{}r@{~}c@{~}c@{}c@{}}
     \tableline
        ID & Frequency\,[$\mu$Hz] & Period\,[s] & amplitude\,[mmi]\\
        \hline\\[-1.8ex]
        f1 & 7311.7522$\pm$0.0008& 136.76612$\pm$0.00016 & 7.9$\pm$0.3 \\
        f2 & 5902.527~~$\pm$0.003~~~& 169.4190~~$\pm$0.0008~~ & 2.4$\pm$0.3 \\
       \tableline
    \end{tabular}
\end{table}
%%%%%%%%%%%%%%%%%%%%%%%%%%%%%%%%%%%%%%%%%%%%%%%%%%%%%%%%%%%%%%%%%%%%%%%%%%%%%%%%%%%%%%%%%%
\begin{table}[b]
\small
\caption{The two main frequencies in {HS\,0702+6043} as obtained from the 2005 Calar Alto data set} %% no full stop at the end of caption
\label{tbl:hs0702frequencies}
    \begin{tabular}{@{}r@{~}c@{~}c@{}c@{}}
     \tableline
        ID & Frequency\,[$\mu$Hz] & Period\,[s] & amplitude\,[mmi]\\
        \hline\\[-1.8ex]
        f1 & 2753.9618$\pm$0.0047& 363.1132$\pm$0.006~   &27.2$\pm$0.2 \\
        f2 & 2606.0110~$\pm$0.0241~~& 383.7282$\pm$0.0035 & 5.3$\pm$0.2 \\
       \tableline
    \end{tabular}
\end{table}
%
%%%%%%%%%%%%%%%%%%%%%%%%%%%%%%%%%%%%%%%%%%%%%%%%%%%%%%%%%%%%%%%%%%%%%%%%%%%%%%%%%%%%%%%%%%

We note that the observations for \object{HS\,0702+6043} comprise a very
large and very high-quality data set obtained by Green \textit{et al.}
in 2007/2008 (''Mt.\ Bigelow data set'') primarily for asteroseismological purposes, which is
presented in detail by \citet{2010ApSS.F} elsewhere in this
issue. We include and examine it here exclusively under the
aspect of long-term changes.

\subsection{Data reduction}\label{ss:reduction}

Light curves were extracted from (bias-, dark- and
flat-field-corrected) CCD frames with aperture photometry packages,
usually directly by the observers. Relative light curves were obtained
using a prescribed set of reference stars for each target. We then
applied an extinction correction to the relative light curves using
low-order polynomials. The timestamps of observations are given at
mid-exposure, and were barycentrically corrected as well as corrected
for leap seconds. The BJD is stripped of the leading 24 and
therefore appears as ''Reduced'' Julian Date in the figures (BRJD).
The normalised relative intensities were given weights
$1/\sigma_{\textrm{res}}^2$ 
%$\displaystyle\frac{1}{\sigma_{\textrm{res}}^2}$ 
according to the local residual intensity scatter.
The subsequent analysis of the weighted light curves was done using
\texttt{Period04} by \citet{2005CoAst.146...53L}. The errors for any
frequencies, amplitudes, phases derived for various subsets of the
data were calculated according to the prescription by
\citet{1999DSSN...13...28M}.

%%%%%%%%%%%%%%%%%%%%%%%%%%%%%%%%%%%%%%%%%%%%%%%%%%%%%%%%%%%%%%%%%%%%%%%%%%%%%%%%%%%%%%%%%%

\begin{figure}[tb]
\includegraphics[height=\columnwidth,angle=90]{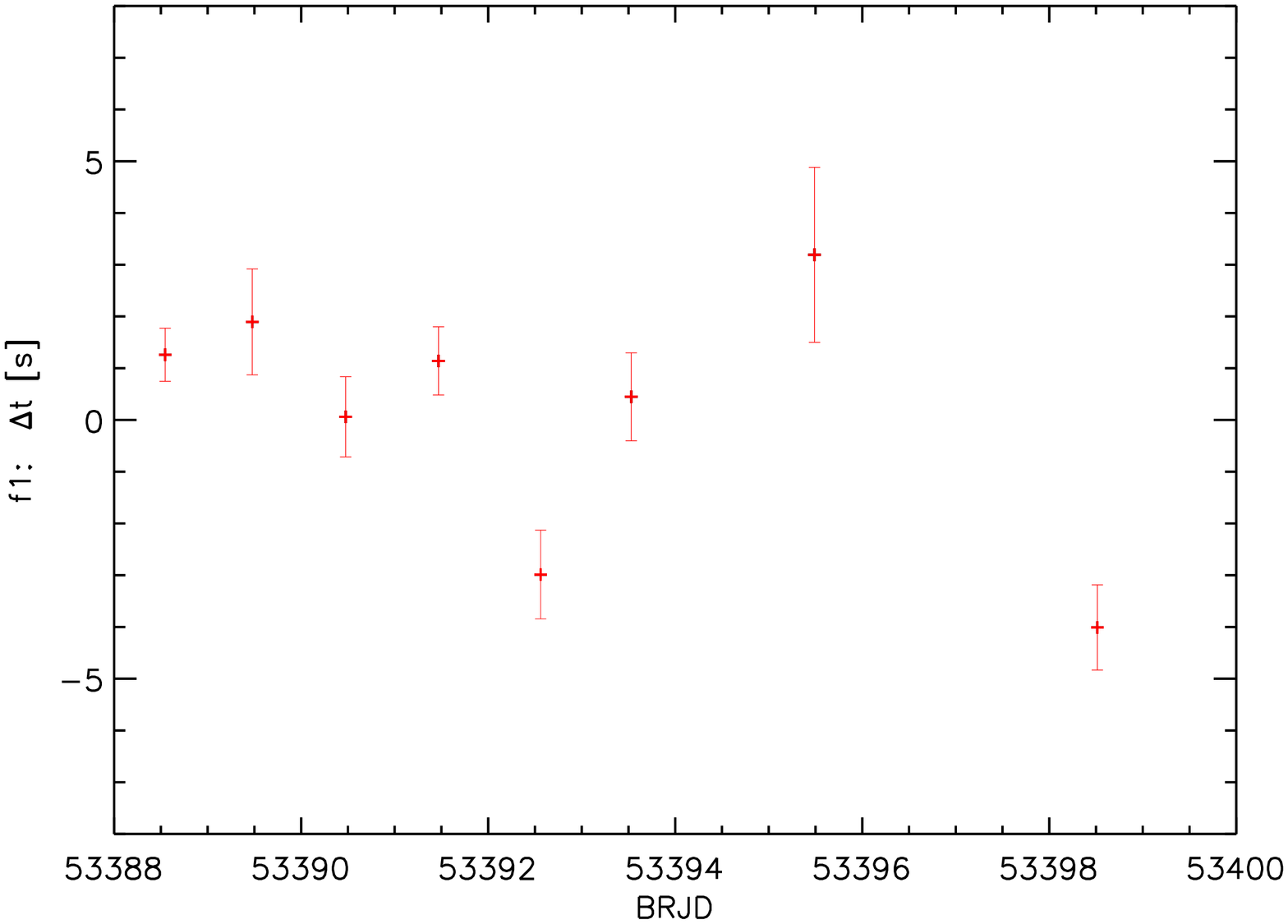}
\caption{8 nights of Calar Alto data on {HS\,0702+6043}: O$-$C diagram for f1} %% no full stop at the end of caption
\label{fig:hs0702caocf1}
\end{figure}

\begin{figure}[tb]
\includegraphics[height=\columnwidth,angle=90]{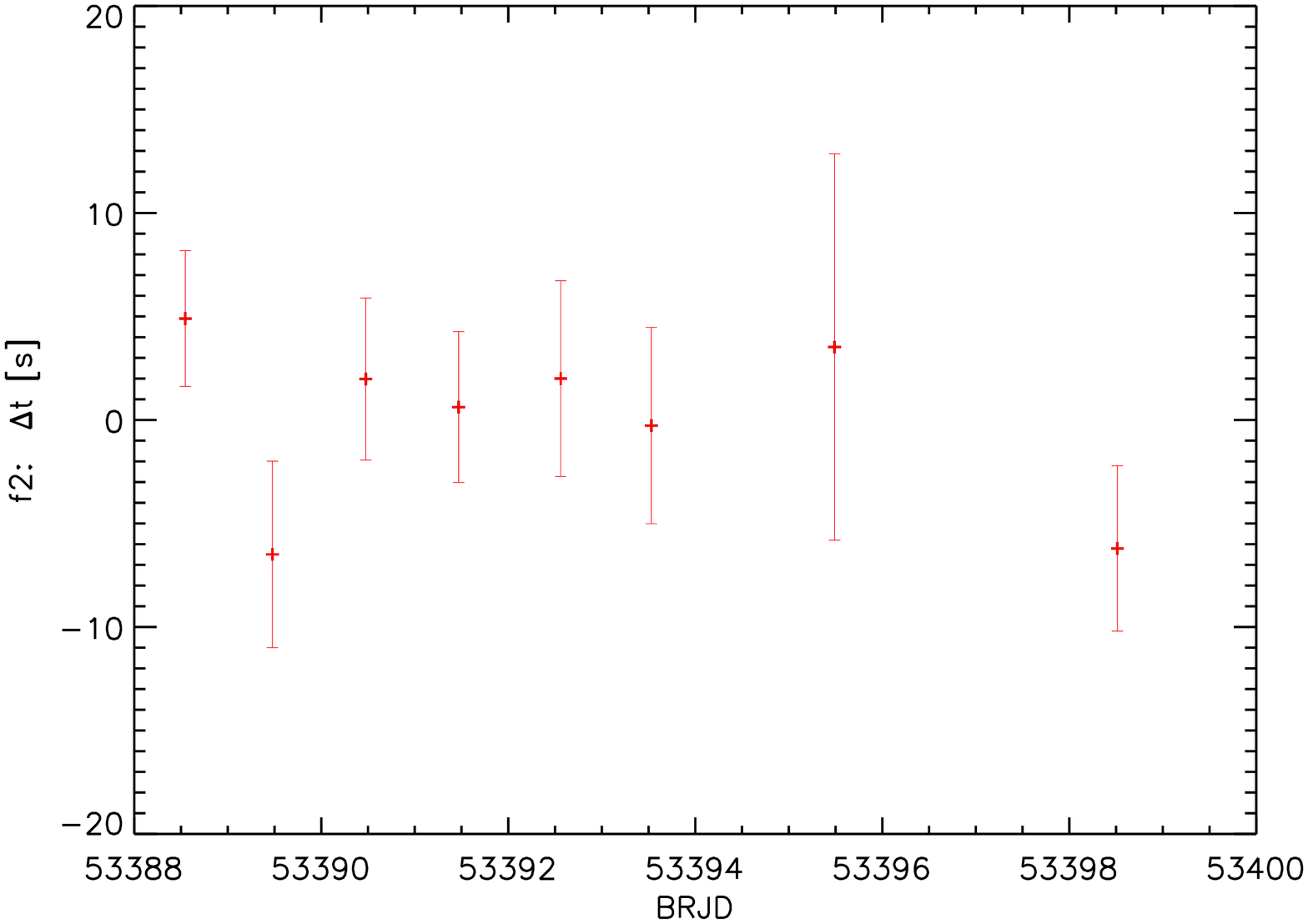}
\caption{8 nights of Calar Alto data on {HS\,0702+6043}: O$-$C diagram for f2} %% no full stop at the end of caption
\label{fig:hs0702caocf2}
\end{figure}

\begin{figure}[tb]
\includegraphics[height=\columnwidth,angle=90]{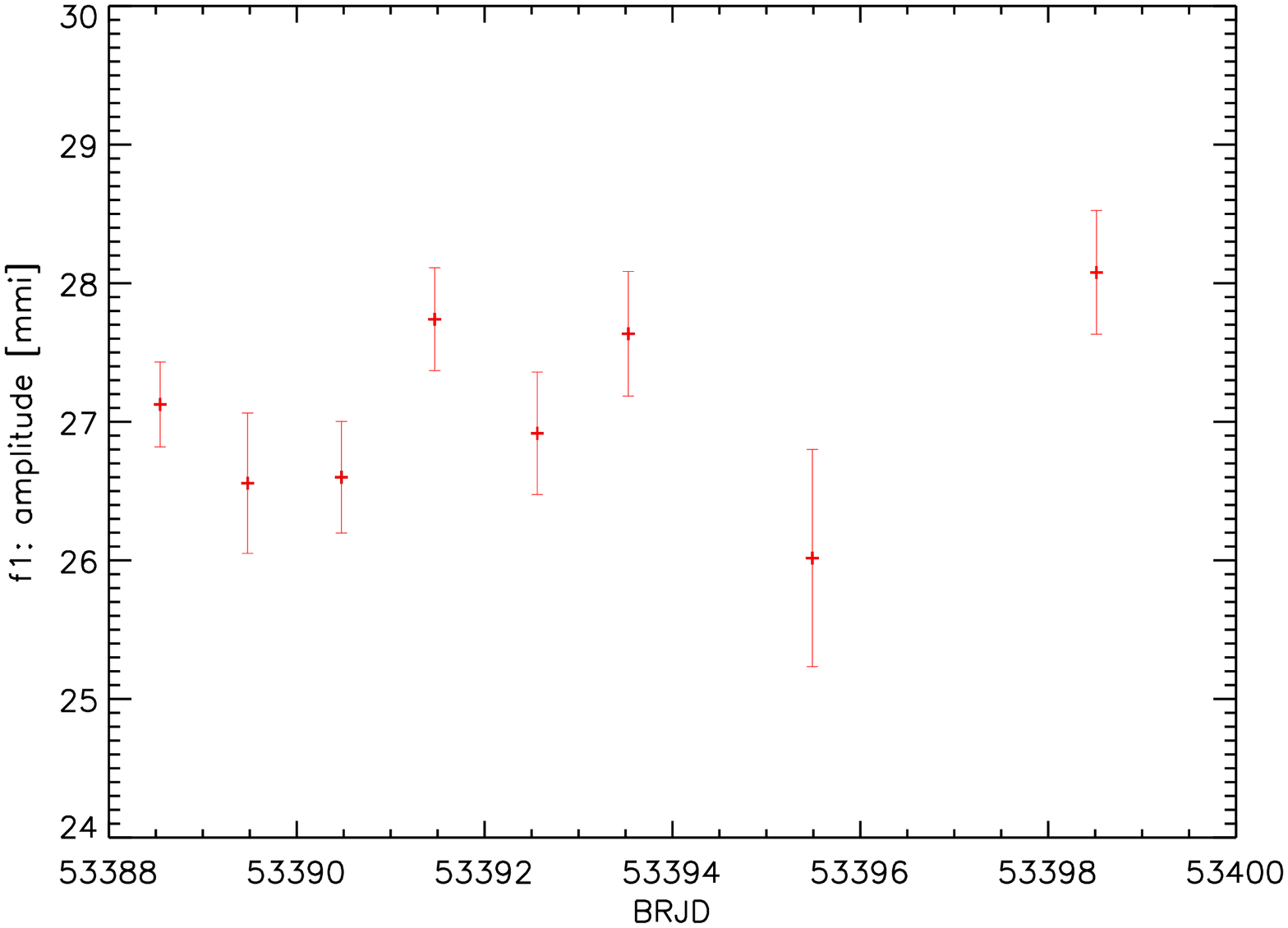}
\caption{8 nights of Calar Alto data on {HS\,0702+6043}: amplitude variation for f1} %% no full stop at the end of caption
\label{fig:hs0702caampf1}
\end{figure}

\begin{figure}[tb]
\includegraphics[height=\columnwidth,angle=90]{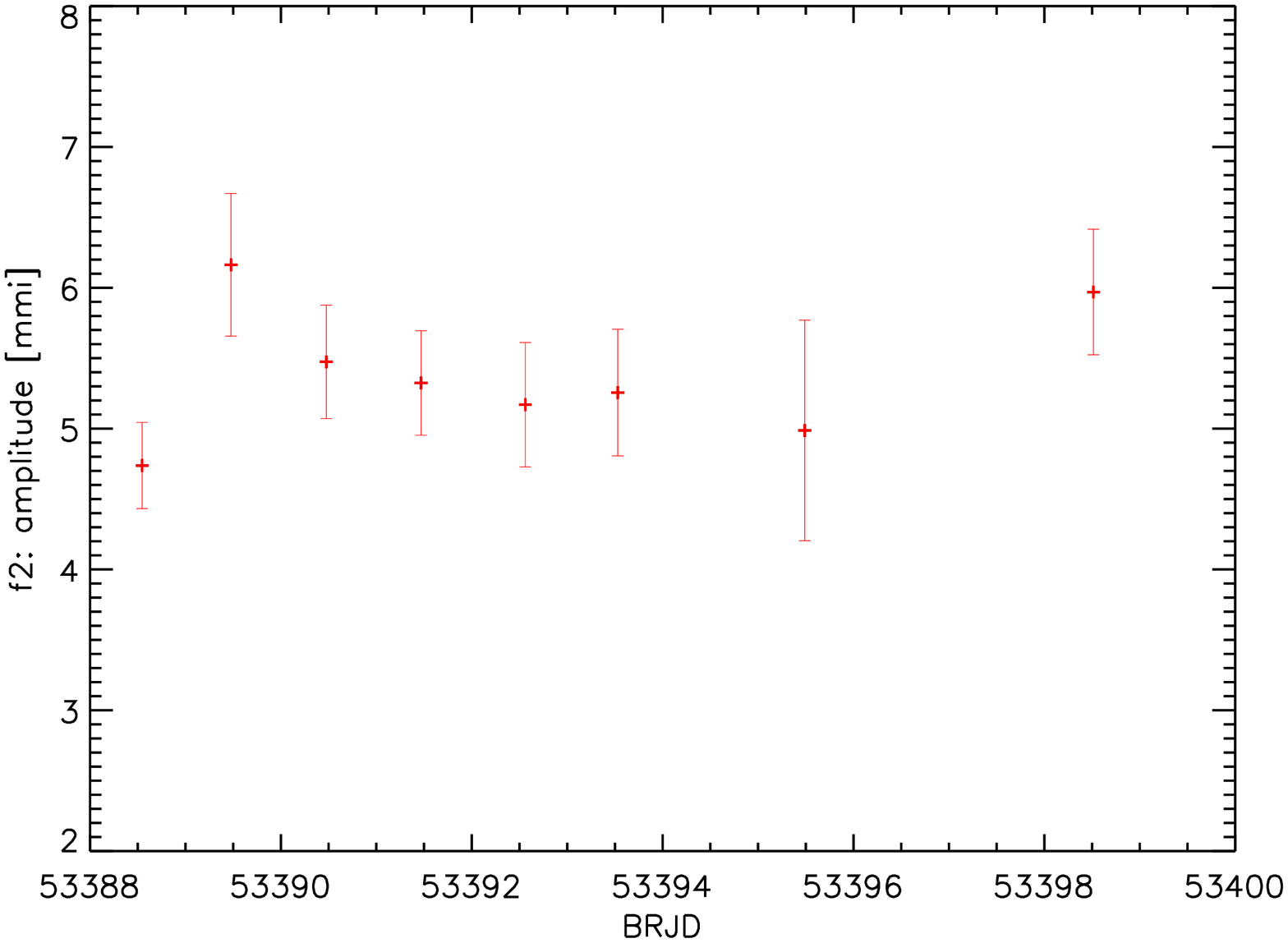}
\caption{8 nights of Calar Alto data on {HS\,0702+6043}: amplitude variation for f2} %% no full stop at the end of caption
\label{fig:hs0702caampf2}
\end{figure}

%%%%%%%%%%%%%%%%%%%%%%%%%%%%%%%%%%%%%%%%%%%%%%%%%%%%%%%%%%%%%%%%%%%%%%%%%%%%%%%%%%%%%%%%%%

\section{Coherence}\label{s:coherence}
\subsection{Photometric~monitoring~of HS0444+0458}\label{ss:hs0444}

We use the six-night data set from TNG in August 2008, which is the
longest individual observing run and the one with the highest S/N
available for \object{HS\,0444+0458}, in order to examine the
short-term stability of the two main pulsations in this star.

The frequencies used in this exercise were derived from the full data
set, the values are as listed in Table~\ref{tbl:hs0444frequencies}.
Keeping these frequencies fixed, we re-derived the best mean amplitude
for the overall TNG data set, as well as the mean phase.  Keeping both
the frequencies and the amplitude fixed, we then turned to the
individual nights to re-determine the phasings. The differences
between the mean phase and the nightly phasings were converted into
time lags, and the errors propagated from the error of the mean phase
and the errors of the nightly phasings. The results for the
frequencies f1 and f2 are shown in Fig.~\ref{fig:hs0444tngocf1} and
Fig.~\ref{fig:hs0444tngocf2}. 

There are no serious discontinuities in
the phases of the pulsations that would have to be attributed to
instabilities in the pulsational behaviour of \object{HS\,0444+0458};
all deviations are consistent with the uncertainties expected from the
measurement uncertainties alone (short individual runs equal
relatively large uncertainties in the phasing compared to phasings
derived from an ensemble of several nights, as used for the
construction of long-term O$-$C diagrams).  

The reasonably flat O$-$C diagrams therefore show that
\object{HS\,0444+0458}'s pulsations are coherent on this time
scale. Extrapolating this result at the given overall frequency
precision, the cycle count remains reliable during the 2-3 month gaps
in between the data sets obtained for \object{HS\,0444+0458}.

Similarly, Fig.~\ref{fig:hs0444tngampf1} and
Fig.~\ref{fig:hs0444tngampf2} show the degree of variation in the
amplitudes for f1 and f2 from night to night during the TNG run.

%%%%%%%%%%%%%%%%%%%%%%%%%%%%%%%%%%%%%%%%%%%%%%%%%%%%%%%%%%%%%%%%%%%%%%%%%%%%%%%%%%%%%%%%%%
\begin{figure}[tb]
\includegraphics[height=\columnwidth,angle=90]{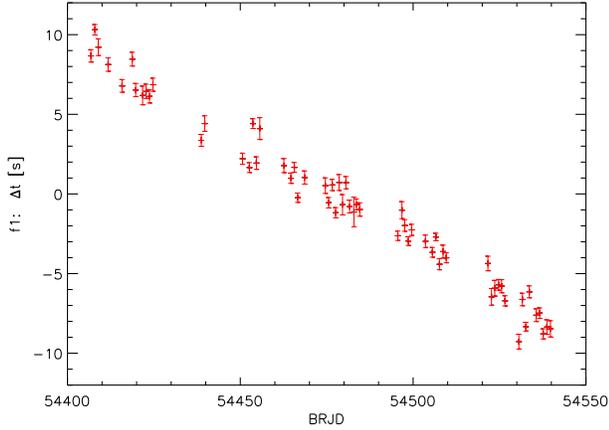}
\caption{60 nights of Mt Bigelow data on {HS\,0702+6043}: O$-$C diagram for f1} %% no full stop at the end of caption
\label{fig:hs0702mtbocf1}
\end{figure}

\begin{figure}[tb]
\includegraphics[height=\columnwidth,angle=90]{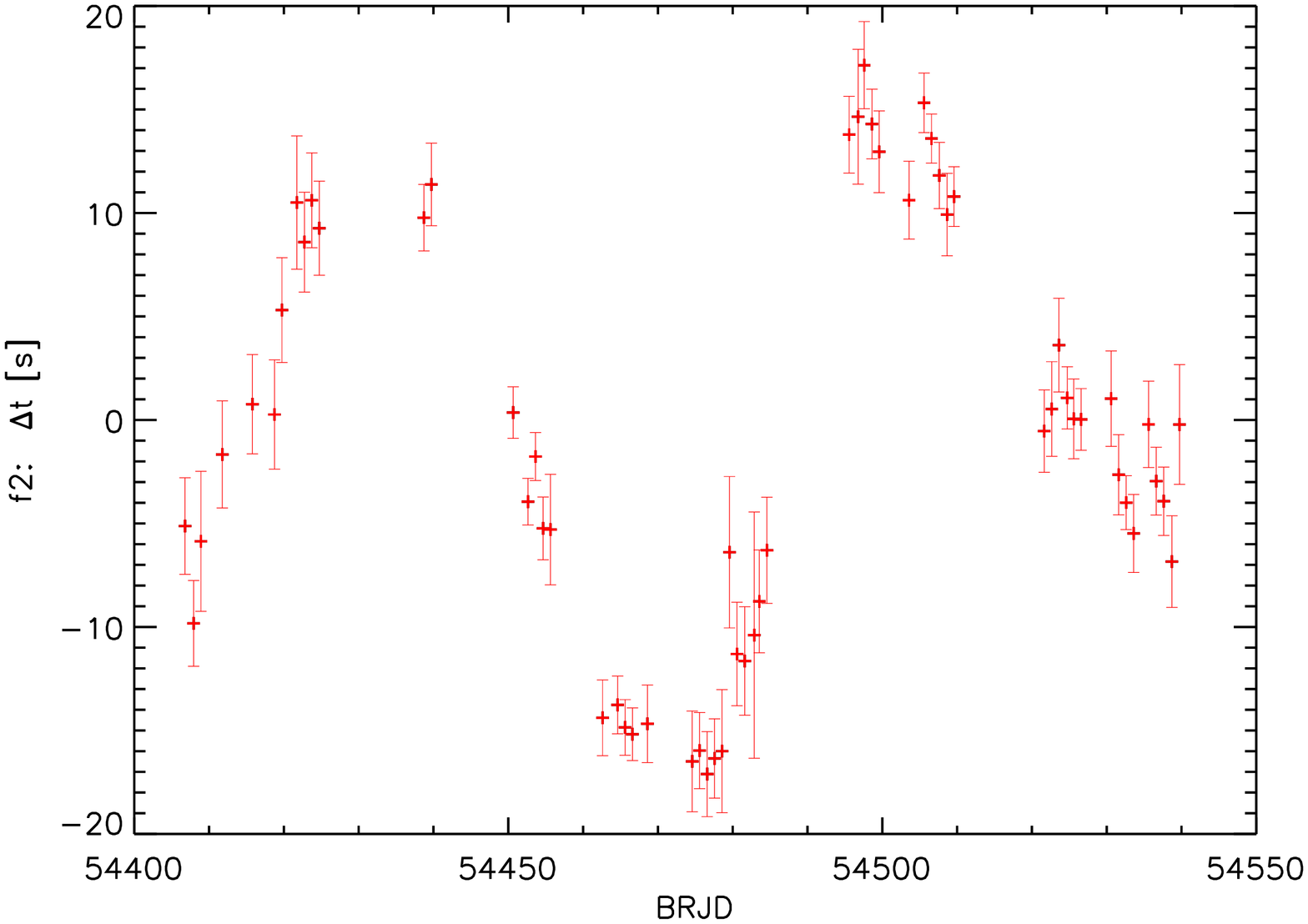}
\caption{60 nights of Mt Bigelow data on {HS\,0702+6043}: O$-$C diagram for f2} %% no full stop at the end of caption
\label{fig:hs0702mtbocf2}
\end{figure}

\begin{figure}[tb]
\includegraphics[height=\columnwidth,angle=90]{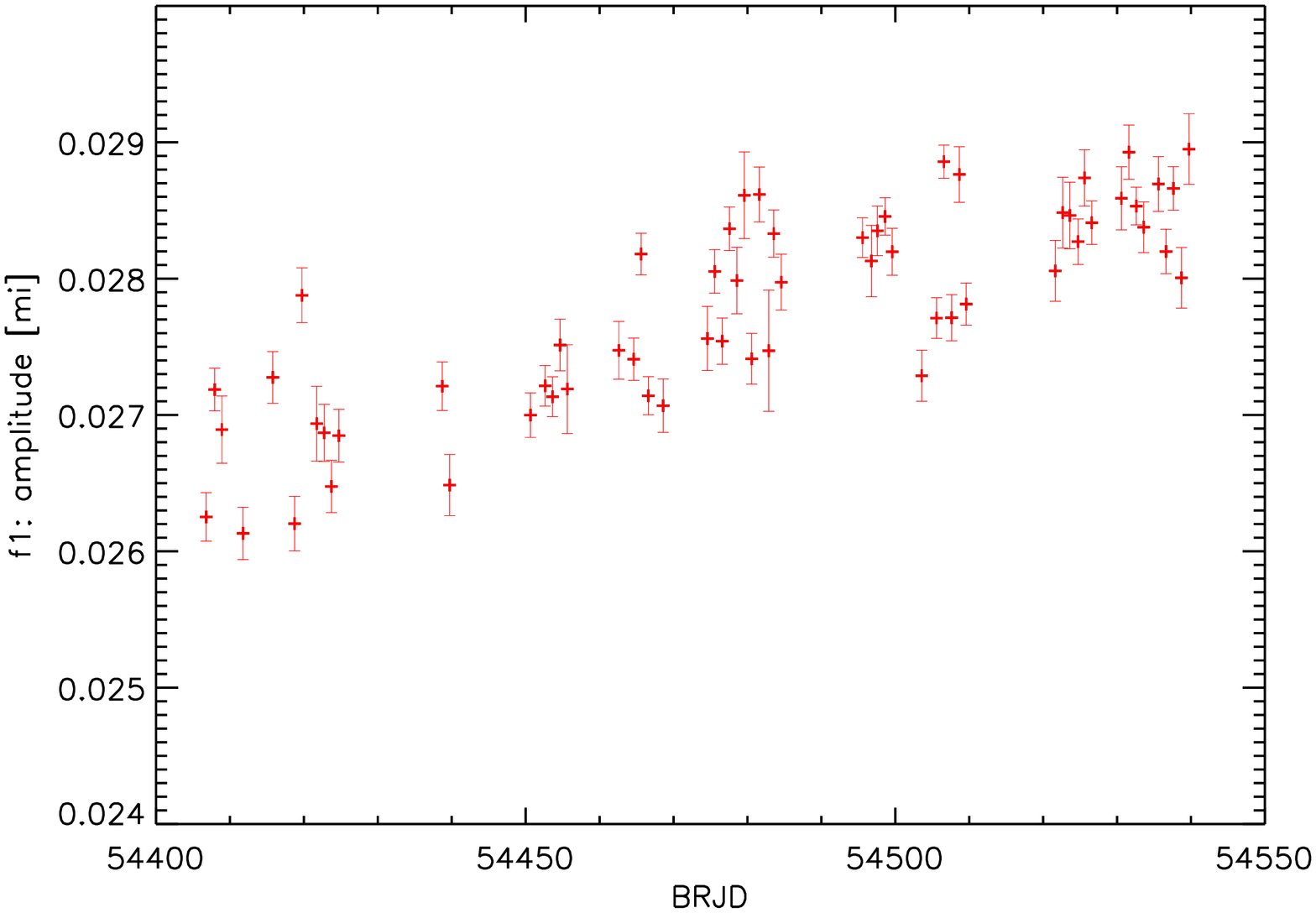}
\caption{60 nights of Mt Bigelow data on {HS\,0702+6043}: amplitude variation for f1} %% no full stop at the end of caption
\label{fig:hs0702mtbampf1}
\end{figure}

\begin{figure}[tb]
\includegraphics[height=\columnwidth,angle=90]{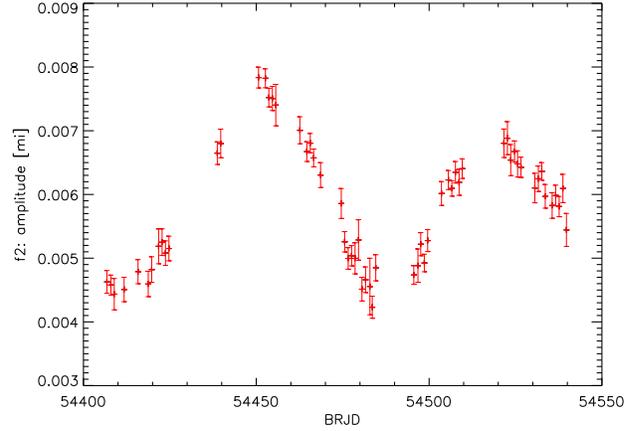}
\caption{60 nights of Mt Bigelow data on {HS\,0702+6043}: amplitude variation for f2} %% no full stop at the end of caption
\label{fig:hs0702mtbampf2}
\end{figure}

%%%%%%%%%%%%%%%%%%%%%%%%%%%%%%%%%%%%%%%%%%%%%%%%%%%%%%%%%%%%%%%%%%%%%%%%%%%%%%%%%%%%%%%%%%%

\subsection{Photometric monitoring of HS\,0702+6043}\label{ss:hs0702}

The coherence and amplitude stability in \object{HS\,0702+6043} were
tested in two independent different data sets. 

The first data set consists of eight nights of data obtained at Calar
Alto in 2005. The two main rapid pulsations were recovered at the
values given in Table~\ref{tbl:hs0702frequencies} (compare also
\citealt{2008ASPC..392..339L}). They were used in the same way as in
the previous section to construct O$-$C diagrams
(Fig.~\ref{fig:hs0702caocf1} and Fig.~\ref{fig:hs0702caocf2}) as well
as to investigate the amplitude stability
(Fig.~\ref{fig:hs0702caampf1} and Fig.~\ref{fig:hs0702caampf2}) for f1
and f2 in \object{HS\,0702+6043}. 

Despite the longer data set and
higher S/N (due to both a brighter target and higher pulsation
amplitudes) and hence smaller error bars, the phases in
\object{HS\,0702+6043} appear more scattered than in\linebreak
\object{HS\,0444+0458}, and the amplitude in particular for the main
frequency f1 significantly less stable.

The second data set comprises the core 60 nights of the Mt.\ Bigelow
run (see \citealt{2010ApSS.F} for the frequency analysis). The most
immediate and notable result from this data set is that
\object{HS\,0702+6043} shows a lot more frequencies both in the p- as
well as in the g-mode domain than previously published.

We treated the data independently as above to obtain near-continuous O$-$C
diagrams spanning more than 100 days for f1 and f2
(Fig.~\ref{fig:hs0702mtbocf1} and Fig.~\ref{fig:hs0702mtbocf2}) as
well as diagrams of the evolution of the corresponding pulsation amplitudes
(Fig.~\ref{fig:hs0702mtbampf1} and Fig.~\ref{fig:hs0702mtbampf2})
during this time. The results for the amplitude variation confirm  
those obtained by \citet{2010ApSS.F}.

All of these figures reveal a rather complex behaviour that needs to
be discussed in detail. The O$-$C diagram for f1 shows a significant
trend which looks more or less linear (which is also seen in the
associated amplitude). Normally this would be a telltale sign for a
mean frequency chosen at a slightly wrong value, which can be
corrected to subsequently get rid of the trend. However, all attempts
to ''improve'' the frequency have so far proven unsuccessful to
eliminate this trend (a result consistent with the mean solution for
the full 2004-2009 data), so it must represent a higher-order effect.

The O$-$C diagram for f2 shows an even more complex behaviour, with
roughly oscillatory shape (with a very similar evolution in the
amplitude), indicative of beating. With less than two full ''beating
cycles'' covered even by this very extensive data set, however, the
hypothetical multiplet structure potentially generating this beating
still cannot be adequately resolved.

It is worth noting that the O$-$C diagrams and amplitude variation
results do not change when increasingly more frequencies are included
simultaneously in the analysis (up to 18 frequencies tested); this
does not include, however, hypothetical additional very close
frequencies within the frequency resolution of frequencies already
considered.

It would therefore be tempting to claim that this data set reveals
first indications for what the \mbox{EXOTIME} program is, after all,
looking for: drifts or even periodic residuals in the O$-$C diagrams
that can only be explained by the influence of substellar
companions. However, \emph{at most one} of the trends in the O$-$C
diagrams could be attributed to one or more unseen companion, as the
light travel time would have to change in exactly the same way and
with exactly the same amplitude for all pulsations, independently of
their other properties. Such a similarity in the patterns of the O$-$C
diagrams for different pulsations has however not been uncovered in
this analysis so far, and all remaining irregularities would
\emph{still} have to be explained differently. In the meantime, it
remains unclear if external effects (including unresolved close
frequencies) causes these features or if the pulsations truly drift
very slowly in amplitude, phase or period.

Finally, the fact that the O$-$C diagrams for the much shorter Calar
Alto data alone does not look so bad (although not as good as that for
\object{HS\,0444+0458}) by itself, while the O$-$C diagrams for the
much longer \emph{continuous} Mt.\ Bigelow data set proves the star's
pulsations to be much more complicated, reminds us that all
conclusions derived from shorter runs should be interpreted
with a fair amount of caution.

\subsection{Follow-up observations of HS\,2201+2610}\label{ss:hs2201}
%\subsubsection{}%\label{sss:?}

The O$-$C diagrams, as well as pulsational amplitudes, derived for
\object{HS\,2201+2610} have been published elsewhere.  An extension of
the diagrams for f1 and f2 of this star is work in progress, with
significant amounts of new data available by now. In addition to the
continuing long-term photometric monitoring data in one filter,
multi-colour data and time-resolved spectra have been obtained.
\citet{2009CoAst.159...91S} sketch an idea of how to combine such
observations to get an estimate of the orbital inclination $i$ for the
companion, which would allow to transform the current $m_2\sin{i}$
measurement into a true mass $m_2$. We note that efforts in this
direction have however only indicated so far that a)~\object{HS\,2201+2610}
is probably a slow rotator and that it b)~has very small pulsational radial
velocity amplitudes.

\section{Long-term O$-$C diagrams}\label{s:longterm}

While the long-term O$-$C diagrams for \object{HS\,2201+2610} are
available elsewhere, their construction for \linebreak
\object{HS\,0702+6043} is work in progress, and is currently still
hampered by the difficulties detailed in section~\ref{ss:hs0702}.

We therefore focus on presenting the preliminary O$-$C diagrams for
\object{HS\,0444+0458} here. The overall coverage (a total of 8 month
of data) is much shorter here in comparison to \object{HS\,2201+2610}
and \object{HS\,0702+6043}, so that obviously no results for $\dot{P}$
or even more complex residuals can be presented yet. 

\begin{figure}[tb]
\includegraphics[height=\columnwidth,angle=90]{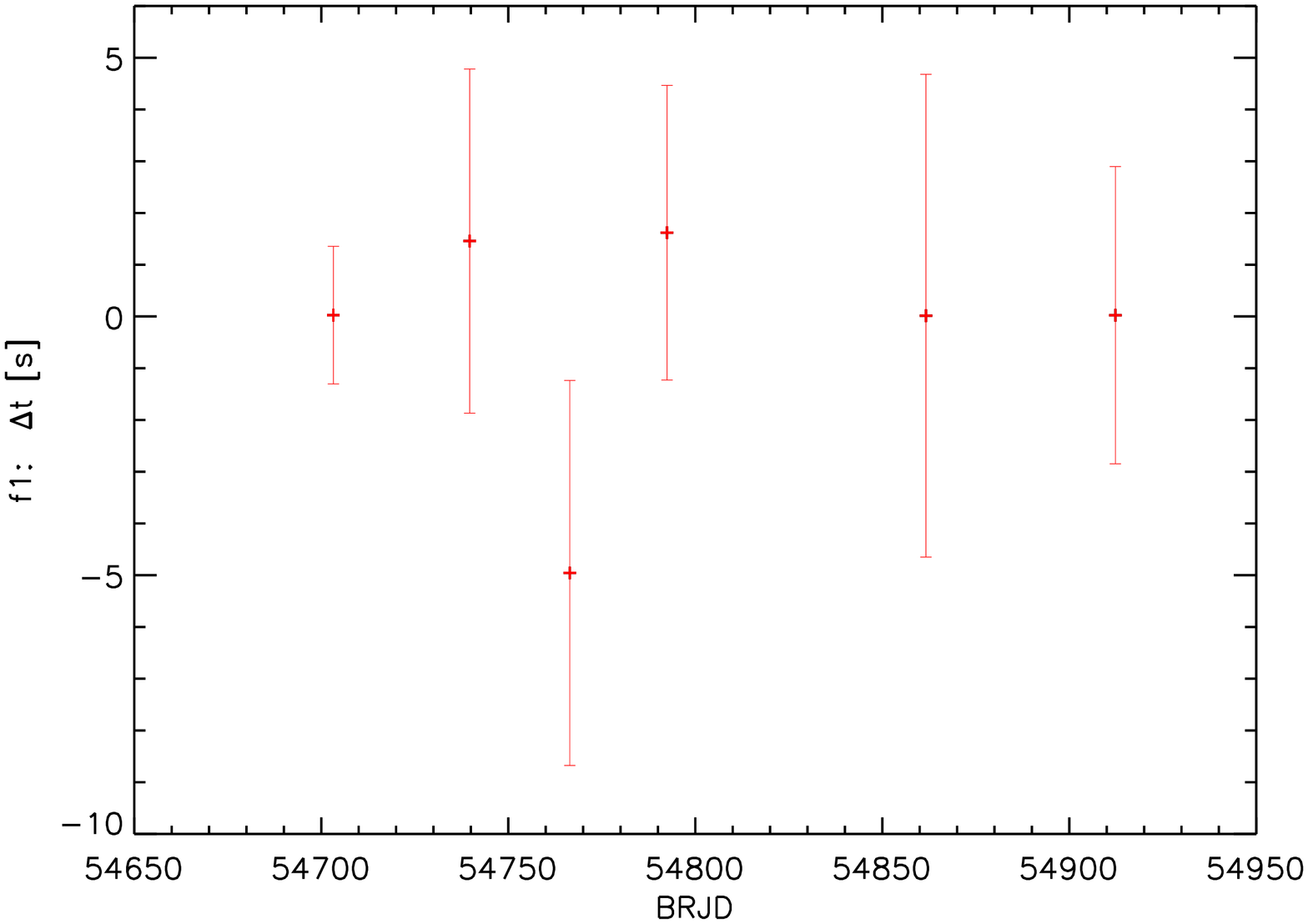}
\caption{All data on {HS\,0444+0458}: O$-$C diagram for f1} %% no full stop at the end of caption
\label{fig:hs0444ocf1}
\end{figure}

\begin{figure}[tb]
\includegraphics[height=\columnwidth,angle=90]{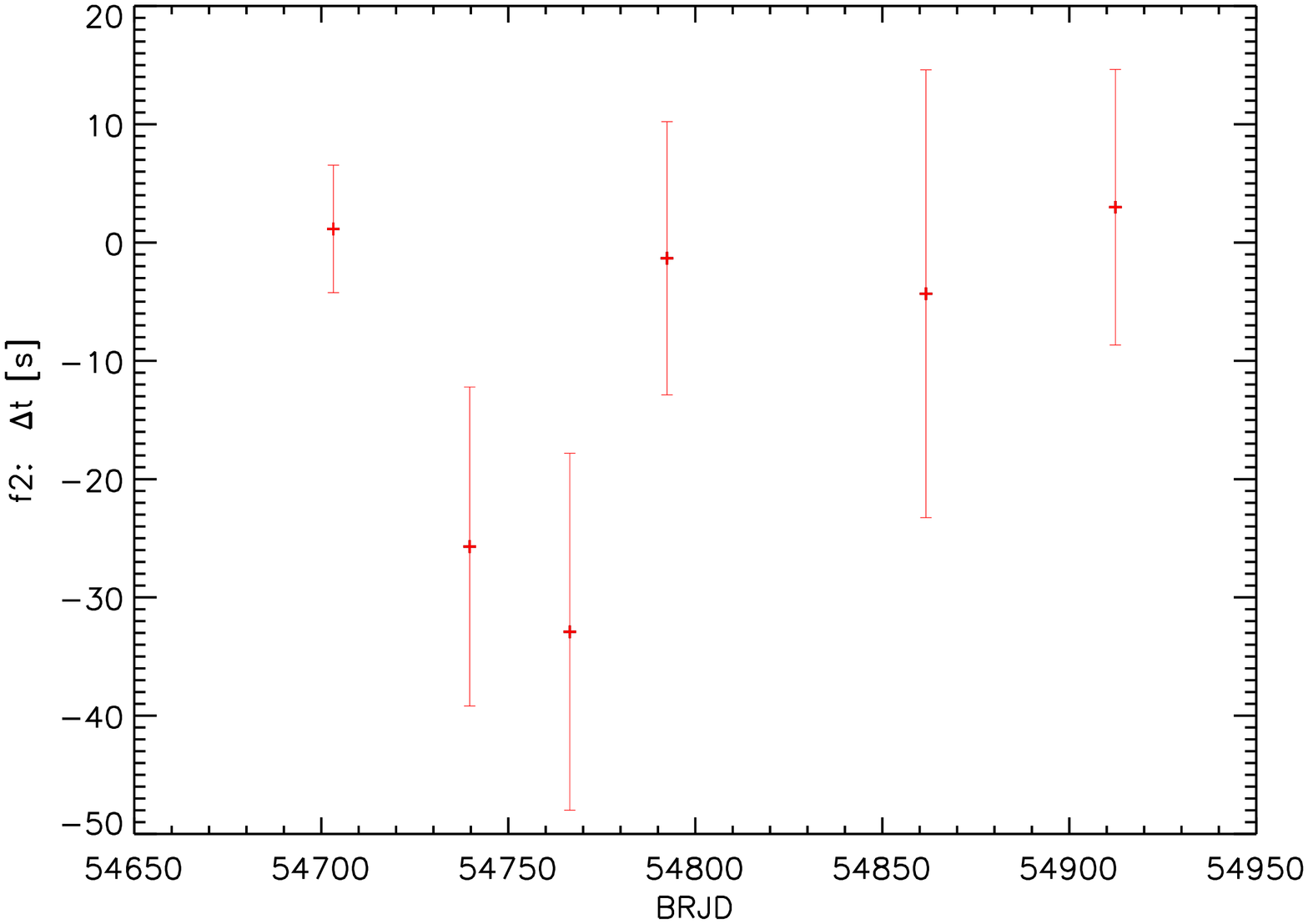}
\caption{All data on {HS\,0444+0458}: O$-$C diagram for f2} %% no full stop at the end of caption
\label{fig:hs0444ocf2}
\end{figure}

The very important preliminary result for this star, however, is that
the O$-$C diagrams for both its frequencies f1 and f2 look fairly flat
within the error bars over the full 8 month period, as can be seen in
Fig.~\ref{fig:hs0444ocf1} and Fig.~\ref{fig:hs0444ocf2}. This means
that \object{HS\,0444+0458} is very well suited as a EXOTIME target
and definitely worth further follow-up.

%%%%%%%%%%%%%%%%%%%%%%%%%%%%%%%%%%%%%%%%%%%%%%%%%%%%%%%%%%%%%%%%%%%%%%%%%%%%%%%%%%%%%%%%%%%

%%%%%%%%%%%%%%%%%%%%%%%%%%%%%%%%%%%%%%%%%%%%%%%%%%%%%%%%%%%%%%%%%%%%%%%%%%%%%%%%%%%%%%%%%%%

\section{Summary}\label{s:summary}

The main result for \object{HS\,0444+0458} so far is that the O$-$C
diagram for its main pulsation frequency f1 is very well-behaved.
The O$-$C diagram for f2 is also approximately flat, but with
much larger errors and hence less significance due to the lower S/N.
We conclude that the degree of coherence of the pulsations in \object{HS\,0444+0458}
confirms it as a good target for long-term monitoring. The next few years will show
if this reveals a measurable $\dot{P}$ for the star and/or the signature of an
orbiting (planetary) companion.

For \object{HS\,0702+6043}, previously thought to show a relatively
simple spectrum in the p-mode domain, it turns out
(\citealt{2010ApSS.F}, and the analysis in this work) that a
significant amount of work must first be put into developing a very
good understanding of the pulsational behaviour before these
pulsations can be used as a tool to derive $\dot{P}$ or even search
for companions.

\section{Discussion}\label{s:discussion}

The increasing number of detections of low-mass companions
\citep{2007Natur.449..189S,2009AJ....137.3181L,2009ApJ...695L.163Q,2009ApJ...702L..96G,2010ApSS.G}
may indicate the existence of a previously undiscovered population of
substellar and planetary bodies around subdwarf B stars.  This has
renewed interest in the question, initially raised by
\cite{1998AJ....116.1308S}, whether planets might be relevant for the
formation of subdwarf B stars. 

%Let's assume here for shortness that the context is well known:
%  existence of observed binaries vs. apparently single
%  mass spectrum of companions
%  associated formation scenarios -> sdB+dM, sdB+WD; NS? BH?

To discuss the possible influence of a substellar companion on the
primary star's evolution on the red giant branch, the above systems
must be considered individually.  In the case of
\object{HS\,2201+2610} \citep{2007Natur.449..189S}, for instance, the
detected planet is too far out to have strongly influenced the
envelope ejection process.
On the other hand, when a star forms planets, it often forms full
planetary systems with several members, so that a remaining detectable
far-out planet may serve as a tracer of previously present closer-in
planets, possibly destroyed during the RGB phase. In this sense, the
detection of \object{V391 Pegasi b} could be interpreted as backing up
the suggestion that planets can indeed enhance the mass loss on the
RGB in such a way that the subsequent formation of a subdwarf B star
is favoured \citep{2010ApSS.S}. The detection of \object{HD 149382 b}
\citep{2009ApJ...702L..96G,2010ApSS.G} may constitute more direct
evidence, with the actual object having undergone the common
envelope phase still present today.

The situation is different for the tertiary bodies in the eclipsing
close binary systems \object{HW Vir} and \object{HS\,0705+6700}
\citep{2009AJ....137.3181L,2009ApJ...695L.163Q}. The ''problem'' of
''missing'' companions that would be needed to explain the formation
of the primary subdwarf B star through one of the accepted binary
evolution channels has never existed here, as suitable secondary
bodies were already known.

The detections of low-mass objects around apparently single sdB stars
are therefore the more relevant ones when it comes to the discussion
whether such companions can be efficient in forming subdwarf B stars.
In particular, the amount of angular moment transferred and hence the
modification of the ejection efficiency during a common envelope phase
are a matter of debate among theorists.
%%% (which) Podsiadlowski reference ???
In alternative scenarios, the existence of planets around subdwarf B
stars is explained by formation processes leading to second generation
planets.

The EXOTIME program has the potential to add to the empirical data
available on which such a debate must ultimately be based.

\begin{comment}
8 pages for talks (minimum 4 pages!)
4 pages per poster
SUM 8+4+4= max 16 pages
\end{comment}

%%%%%%%%%%%%%%%%%%%%%%%%%%%%%%%%%%%%%%%%%%%%%%%%%%%%%%%%%%%%%%%%%%%%%%%%%%

\begin{table*}[tb]
\small
\caption{Photometric data archive for HS0444+0458} %% no full stop at the end of caption
\label{tbl:hs0444}
\begin{tabular}{cccrrrlllll}
\tableline  %% rule at top
Year & Month & Day & L [h] & N & $\sigma_{\textrm{res}}$ & Exp [s] & Site\tablenotemark{a} &
Inst. & Filter & Observer\tablenotemark{b}\\
\tableline
2008 & Aug & 19 & 0.5 & 107 & 0.0069 & 5 & TNG~3.6m & DOLORES & B & RS, SL\\
     &     & 21 & 1.0 & 237 & 0.0076 & 5 & TNG~3.6m & DOLORES & B & RS, SL\\
     &     & 24 & 0.7 & 173 & 0.0060 & 5 & TNG~3.6m & DOLORES & B & RS, SL\\
     &     & 26 & 1.0 & 224 & 0.0051 & 5 & TNG~3.6m & DOLORES & B & RS, SL\\
     &     & 27 & 2.1 & 357 & 0.0040 & 10 & TNG~3.6m & DOLORES & B & RS, SL\\
     &     & 28 & 1.7 & 230 & 0.0042 & 10 & TNG~3.6m & DOLORES & B & RS, SL\\
     & Sep & 29 & 2.2 & 171 & 0.0089 & 10 & CA~2.2m & CAFOS & B & service (FH)\\
     & Oct & 24 & 2.8 & 300 & 0.0191 & 12 & LOAO~1.0m & 2k CCD & B & SLK\\
     &     & 26 & 4.9 & 588 & 0.0232 & 12 & LOAO~1.0m & 2k CCD & B & SLK\\
     &     & 29 & 1.2 & 163 & 0.0175 & 12 & LOAO~1.0m & 2k CCD & B & SLK\\
     & Nov & 21 & 3.2 & 263 & 0.0092 & 10 & CA~2.2m & CAFOS & B & service (MA)\\
\tableline
2009 & Jan & 29 & 2.6 & 275 & 0.0239 & 20 & LOAO~1.0m & 2k CCD & B & SLK\\
     &     & 31 & 4.7 & 497 & 0.0275 & 20 & LOAO~1.0m & 2k CCD & B & SLK\\
     & Mar & 20 & 1.0 & 94 & 0.0090 & 10 & CA~2.2m & CAFOS & B & service (MA)\\
     &     & 21 & 1.3 & 135 & 0.0121 & 10 & CA~2.2m & CAFOS & B & service (MA)\\
     &     & 22 & 1.2 & 124 & 0.0099 & 10 & CA~2.2m & CAFOS & B & service (MA)\\
     & Oct & 16 & 1.3 & 195 & 0.0138 & 20 & M/N~1.2m & 1k CCD & B & RL\\
     &     & 18 & 1.6 & 242 & 0.0148 & 20 & M/N~1.2m & 1k CCD & B & RL\\
     &     & 22 & 2.1 & 235 & 0.0102 & 10 & CA~2.2m & CAFOS & B & service (AA)\\
     &     & 23 & 2.9 & 330 & 0.0082 & 10 & CA~2.2m & CAFOS & B & service (MA)\\
     &     & 24 & 2.2 & 244 & 0.0075 & 10 & CA~2.2m & CAFOS & B & service (MA)\\
     &     & 25 & 1.7 & 104 & 0.0061 & 10 & CA~2.2m & CAFOS & B & service (MA)\\
     &     & 26 & 3.4 & 387 & 0.0086 & 10 & CA~2.2m & CAFOS & B & service (MA)\\
\tableline %% rule at bottom
\end{tabular}
% Any table notes must follow the \end{tabular} command. 
\tablenotetext{a}{CA: Calar Alto Observatory, LOAO:
  Mt. Lemmon Optical Astronomy Observatory, M/N:
  MONET/North Telescope, TNG: Telescopio Nazionale Galileo}
\tablenotetext{b}{FH: Felipe Hoyo, MA: Manuel Alises, RL: Ronny Lutz, RS:
  Roberto Silvotti, SLK: Seung-Lee Kim, SL: Silvio Leccia}
%\tablecomments{some more comments here...}
\end{table*}

\begin{table*}[tb]
\small
\caption{Photometric data archive for HS\,0702+6043, part a} %% no full stop at the end of caption
\label{tbl:hs0702a}
\begin{tabular}{cccrrrlllll}
\tableline  %% rule at top
Year & Month & Day & L [h] & N & $\sigma_{\textrm{res}}$ & Exp [s] & Site\tablenotemark{a} &
Inst. & Filter & Observer\tablenotemark{b}\\
\tableline
1999 & Dec & 06 & 5.0 & 943 & 0.0103 & 10 & CA~1.2m & CCD & - & SD, SS\\
     &     & 07 & 1.9 & 414 & 0.0141 & ~\,8 & CA~1.2m & CCD & - & SD, SS\\
     &     & 08 & 1.4 & 241 & 0.0085 & 10 & CA~1.2m & CCD & - & SD, SS\\
\tableline
2000 & Oct & 08 & 0.7 & 248 & 0.0027 & 10 & NOT~2.56m & ALFOSC & 3W & R\O\\
\tableline
2004 & Feb & 04 & 7.3 & 685 & 0.0233 & 30 & Tue~0.8m & ST7 & G & TN \\
     &     & 09 & 5.8 & 244 & 0.0016 & 60 & StB~2.2m & 2k CCD & F555W & EMG\\
     &     & 10 & 6.2 & 264 & 0.0014 & 60 & StB~2.2m & 2k CCD & F555W & EMG\\
\tableline
2005 & Jan & 17-27 & 56.0 & 5981 & 0.0082 & 20, 30 & CA~2.2m & CAFOS & B & SS, TS\\
\tableline
%2007-08 & Nov-Mar &  & 424.0 & 31638 & 0.0027 &  & MtB~1.55m & MONT~4k & S 8612 & EMG\\
2007 & Dec & 18 & 5.9 & 498 & 0.0442 & 40 & Tue~0.8m & SBIG ST-L & B & TN\\
     &     & 19 & 6.7 & 557 & 0.0589 & 40 & Tue~0.8m & SBIG ST-L & B & TN\\
     &     & 20 & 10.0 & 842 & 0.0994 & 40 & Tue~0.8m & SBIG ST-L & B & TN\\
     &     & 21 & 9.3 & 760 & 0.1350 & 40 & Tue~0.8m & SBIG ST-L & B & TN\\
     &     &    & 12.1 & 621 & 0.0200 &  & Goe~0.5m & SBIG ST-L & B & SS, RL, RK\\
\tableline
2007 & Nov &    &    &    &    &    &    &    &    &\\[-1ex]
 \vdots   &  \vdots  &    & 424.0 & 31638 & 0.0027 & 50 & MtB~1.55m & MONT~4k &
 S 8612 & EMG \& coll.\\
2008 & Mar &    &    &    &    &    &    &    &    &\\
\tableline
2008 & Feb & 08 & 7.2 & 601 & 0.0124 & 40 & Tue~0.8m & SBIG ST-L & B & TN\\
     &     & 09 & 7.1 & 590 & 0.0121 & 40 & Tue~0.8m & SBIG ST-L & B & TN\\
     &     &   & 8.9 & 642 & 0.0381 & 35 & Goe~0.5m & SBIG ST-L & B & RL, BB\\
     &     & 10 & 5.8 & 488 & 0.0134 & 40 & Tue~0.8m & SBIG ST-L & B & TN\\
     &     &    & 12.2 & 910 & 0.0324 & 35 & Goe~0.5m & SBIG ST-L & B & RL, MH\\
     &     & 11 & 4.5 & 338 & 0.0439 & 35 & Goe~0.5m & SBIG ST-L & B & RL, MH\\
     &     & 13 & 1.1 & 86 & 0.0749 & 35 & Goe~0.5m & SBIG ST-L & B & RL\\
     &     & 15 & 5.4 & 379 & 0.0552 & 35 & Goe~0.5m & SBIG ST-L & B & RL, SW\\
     &     & 29 & 3.7 & 360 &  & 20 & Loi~1.5m & & B  & service (ADB)\\
     & Mar & 01 & 5.7 & 451 &  & 25 & Loi~1.5m & & B  & service (ADB)\\
     &     & 03 & 5.1 & 429 &  & 25 & Loi~1.5m & & B  & service (ADB)\\
     &     & 05 & 0.6 & 47 & 0.0238 & 35 & Goe~0.5m & SBIG ST-L & B & RL, SS\\
     &     & 13 & 3.7 & 126 & 0.0069 & 30 & Asi~1.8 & Afosc & R & SB, RC\\
     &     &    & 2.3 & 248 & 0.0105 & 30 & Kon~1.0m & PI VA 1300B & B & MP, PP\\
     &     & 15 & 4.6 & 597 & 0.0095 & 25 & Kon~1.0m & PI VA 1300B & B & MP, PP\\
     &     & 18 & 3.7 & 348 & 0.0111 & 35 & Kon~1.0m & PI VA 1300B & B & MP, PP\\
     &     & 19 & 1.1 & 137 & 0.0088 & 25 & Kon~1.0m & PI VA 1300B & B & MP, PP\\
     & May & 11 & 3.0 & 204 & 0.0315 & 40 & Goe~0.5m & SBIG ST-L & B & SW\\
     &     & 12 & 1.2 & 160 & 0.0380 & 40 & Goe~0.5m & SBIG ST-L & B & SW\\
     &     & 13 & 3.2 & 266 & 0.0254 & 40 & Tue~0.8m & SBIG ST-L & B & AH\\
     & Oct & 15 & 1.3 & 113 & 0.0085 & 10 & CA~2.2m & CAFOS & B & service (LM)\\
     &     & 16 & 0.8 & 65  & 0.0101 & 10 & CA~2.2m & CAFOS & B & service (MA)\\
     &     & 17 & 0.5 & 48  & 0.0152 & 10 & CA~2.2m & CAFOS & B & service (MA)\\
     &     & 18 & 1.6 & 143 & 0.0068 & 10 & CA~2.2m & CAFOS & B & service (MA)\\
     &     & 25 & 2.6 & 285 & 0.0089 & 30 & Kon~1.0m & PI VA 1300B & B & ZB\\
     &     &    & 4.4 & 518 & 0.0107 & 15 & LOAO~1.0m & 2k CCD & B & SLK\\
     &     & 28 & 2.9 & 326 & 0.0104 & 15 & LOAO~1.0m & 2k CCD & B & SLK\\
     &     & 30 & 3.9 & 455 & 0.0104 & 15 & LOAO~1.0m & 2k CCD & B & SLK\\
     & Nov & 21 & 3.0 & 254 & 0.0048 & 10 & CA~2.2m & CAFOS & B & service (MA)\\
     &     & 23 & 5.7 & 496 & 0.0064 & 10 & CA~2.2m & CAFOS & B & service (MA)\\ 
     & Dec & 16 & 3.0 & 243 & 0.0068 & 25 & M/N~1.2m & 1k CCD & B & RL, BL\\
     &     & 20 & 2.7 & 202 & 0.0116 & 25 & M/N~1.2m & 1k CCD & B & RL\\
\tableline
2009 & Jan & 06 & 1.6 & 143 & 0.0364 & 30 & Goe~0.5m & SBIG ST-L & B & RL, US\\
     &     & 22 & 3.7 & 334 & 0.0080 & 25 & M/N~1.2m & 1k CCD & B & RL\\
     &     & 28 & 5.4 & 473 & 0.0147 & 20 & LOAO~1.0m & 2k CCD & B & SLK\\
     &     & 30 & 6.6 & 662 & 0.0138 & 20 & LOAO~1.0m & 2k CCD & B & SLK\\
\tableline %% rule at bottom
\end{tabular}
% Any table notes must follow the \end{tabular} command. 
\tablenotetext{a}{see footnote in Table \ref{tbl:hs0702b} for the site abbreviations}
\tablenotetext{b}{see footnote in Table \ref{tbl:hs0702b} for the observer abbreviations}
%\tablecomments{more comments here...}
\end{table*}

\begin{table*}[tb]
\small
\caption{Photometric data archive for HS\,0702+6043, part b} %% no full stop at the end of caption
\label{tbl:hs0702b}
\begin{tabular}{cccrrrlllll}
\tableline  %% rule at top
Year & Month & Day & L [h] & N & $\sigma_{\textrm{res}}$ & Exp [s] & Site\tablenotemark{a} &
Inst. & Filter & Observer\tablenotemark{b}\\
\tableline
2009 & Feb & 18 & 2.4 & 202 & 0.0256 & 30 & Goe~0.5m & SBIG ST-L & B & RL, TOH\\
%     &     & 23-28 &  &  &  &  & TT1~1.5m &  &  & DM, MD, FC\\
     &     & 23 & 3.6 & 304 & 0.0126 & 25 & M/N~1.2m & 1k CCD & B & RL, US, MH\\
     &     & 25 & 6.4 & 564 & 0.0076 & 25 & M/N~1.2m & 1k CCD & B & RL\\
     &     & 26 & 6.5 & 536 & 0.0111 & 25 & M/N~1.2m & 1k CCD & B & RL, LN\\
     &     & 27 & 6.6 & 556 & 0.0083 & 25 & M/N~1.2m & 1k CCD & B & RL\\
     &     &    & 5.7 & 532 & 0.0067 & 15 & Asi~1.8m & Afosc & no & SB, RC\\
     & Mar & 01 & 6.2 & 497 & 0.0090 & 25 & M/N~1.2m & 1k CCD & B & RL\\
     &     & 18 & 3.3 & 279 & 0.0060 & 25 & M/N~1.2m & 1k CCD & B & RL\\
     &     & 19 & 3.6 & 291 & 0.0098 & 25 & M/N~1.2m & 1k CCD & B & RL\\
     &     & 20 & 2.4 & 243 & 0.0058 & 10 & CA~2.2m & CAFOS & B & service (MA)\\
     &     & 21 & 2.4 & 232 & 0.0068 & 10 & CA~2.2m & CAFOS & B & service (MA)\\
     &     &    & 6.0 & 351 & 0.0150 & 30 & Asi~1.8m & Afosc & B & SB, RC\\
     &     & 22 & 2.4 & 245 & 0.0049 & 10 & CA~2.2m & CAFOS & B & service (MA)\\
     &     &    & 4.9 & 279 & 0.0140 & 30 & Asi~1.8m & Afosc & B & SB, RC\\
     &     & 24 & 3.4 & 243 & 0.0092 & 25 & M/N~1.2m & 1k CCD & B & RL\\
     &     & 26 & 2.8 & 251 & 0.0081 & 25 & M/N~1.2m & 1k CCD & B & RL, US\\
     &     &    & 3.9 & 396 & 0.0118 & 20 & LOAO~1.0m & 2k CCD & B & SLK\\
     &     & 27 & 1.3 & 160 & 0.0185 & 20 & LOAO~1.0m & 2k CCD & B & SLK\\
     &     & 28 & 3.9 & 404 & 0.0172 & 20 & LOAO~1.0m & 2k CCD & B & SLK\\
     &     & 29 & 2.9 & 309 & 0.0112 & 20 & LOAO~1.0m & 2k CCD & B & SLK\\
     &     & 30 & 3.8 & 386 & 0.0144 & 20 & LOAO~1.0m & 2k CCD & B & SLK\\
     &     & 31 & 4.0 & 399 & 0.0154 & 20 & LOAO~1.0m & 2k CCD & B & SLK\\
     & Sep & 27 & 3.9 & 406 & 0.0132 & 25 & M/N~1.2m & 1k CCD & B & RL\\
     &     & 28 & 4.0 & 420 & 0.0117 & 25 & M/N~1.2m & 1k CCD & B & RL\\
     & Oct & 16 & 4.0 & 412 & 0.0099 & 25 & M/N~1.2m & 1k CCD & B & RL\\
     &     & 18 & 0.9 & 100 & 0.0109 & 25 & M/N~1.2m & 1k CCD & B & RL\\
     &     & 19 & 6.1 & 645 & 0.0119 & 25 & M/N~1.2m & 1k CCD & B & RL\\
     &     & 22 & 3.9 & 359 & 0.0076 & 10 & CA~2.2m & CAFOS & B & service (AA)\\
     &     & 23 & 3.6 & 334 & 0.0067 & 10 & CA~2.2m & CAFOS & B & service (MA)\\
     &     & 24 & 3.1 & 289 & 0.0061 & 10 & CA~2.2m & CAFOS & B & service (MA)\\
     &     & 25 & 2.8 & 297 & 0.0082 & 25 & M/N~1.2m & 1k CCD & B & RL\\
     &     &    & 3.4 & 315 & 0.0078 & 10 & CA~2.2m & CAFOS & B & service (MA)\\
     &     & 26 & 3.0 & 279 & 0.0064 & 10 & CA~2.2m & CAFOS & B & service (MA)\\
\tableline %% rule at bottom
\end{tabular}
% Any table notes must follow the \end{tabular} command. 
\tablenotetext{a}{Asi: Asiago 182cm Copernico Telescope, CA: Calar Alto Observatory, 
  Goe: G\"ottingen IAG 50cm Telescope, Kon: Konkoly RCC Telescope, LOAO:
  Mt. Lemmon Optical Astronomy Observatory, Loi: Loiano 152cm Telescope, M/N:
  MONET/North Telescope, MtB: Mt. Bigelow Kuiper Telescope, NOT: Nordic 
  Optical Telescope, StB: Steward Observatory Bok Telescope, Tue:
  T\"ubingen 80cm Telescope}
\tablenotetext{b}{AA: Alberto Aguirre, ADB: Antonio De Blasi, AH: Agnes Hoffmann, BB: Benjamin
  Beeck, BL: Bj\"orn Loeptien, EMG: Elizabeth M.~Green, LM: Luzma Montoya, LN:
  Lisa Nortmann, MA: Manuel Alises, MH: Markus Hundertmark, MP: Margit Papar\'o,
  PP: P\'eter P\'apics, RC: Riccardo Claudi, RK: Renate Kruspe, RL: Ronny Lutz, R\O: Roy \O stensen, SB:
  Serena Benatti, SD: Stefan Dreizler, SLK: Seung-Lee Kim, SS: Sonja Schuh,
  SW: Sascha Werhahn, TN: Thorsten Nagel, TOH: Tim-Oliver Husser, TS:
  Thorsten Stahn, US: Ulf Seemann, ZB: Zs\'ofia Bogn\'ar}
%\tablecomments{comments here...}
\end{table*}

%% Acknowledgements
%
%
\acknowledgments
The authors thankfully acknowledge the contributions by
J.\ Dittmann,
A.G.\ Fay,
J.\ Laird,
C.M.\ Limbach,
J.\ Portouw,
M.\ Revelle,
J.M.\ Sierchio,
S.M.\ Story,
C.\ Stratton,
A.\ Strom,
P.\ Wroblewski
(all Steward Observatory),
who helped obtain the Mt.\ Bigelow data set.
S.S.\ acknowledges funding through the Eberhard-Karls-Universit\"at
T\"ubingen's Teaching Equality program, and also thanks the DAAD for
allocating a travel grant to attend the Forth Meeting on Hot Subdwarfs
and Related Objects where this paper was presented.
The work of R.L.\ on this project is funded by a stipend 
from the IMPRS PhD program at MPS Katlenburg-Lindau.
C.R.L.\ acknowledges an {\em \'Angeles Alvari\~ no} contract of the regional
government {\em Xunta de Galicia}.
Amongst others, the work in this paper is%\\[-4ex]
\begin{itemize}
\item[-] based on observations collected at the Centro Astron{\'o}mico Hispano Alem{\'a}n
(CAHA) at Calar Alto, operated jointly by the Max-Planck Institut f\"ur
Astronomie and the Instituto de Astrof{\'\i}sica de Andaluc{\'\i}a (CSIC).
\item[-] based on observations made with the Italian Telescopio Nazionale
Galileo (TNG) operated on the island of La Palma by the Fundación
Galileo Galilei of the INAF (Istituto Nazionale di Astrofisica) at the
Spanish Observatorio del Roque de los Muchachos of the Instituto de
Astrofisica de Canarias.
\item[-] based on data obtained with the MOnitoring NEtwork of Telescopes (MONET),
funded by the 'Astronomie \& Internet' program of the Alfried Krupp von Bohlen
und Halbach Foundation, Essen, and operated by the Georg-August-Universit\"at
G\"ottingen, the McDonald Observatory of the University of Texas at Austin, and
the South African Astronomical Observatory.\\
\end{itemize}

%% References
%% Please cite all reference entries in the article text using \cite or
%% equivalent command. 

%%%  Using BibTeX  (Name-Year style)
%
%%%\bibliographystyle{spr-mp-nameyear-cnd}  %% BibTeX style
%%%\bibliography{schuh}                %% BibTeX data

%% Non-BibTeX  (Name-Year style)
%
% \begin{thebibliography}{}
% \bibitem[\protect\citeauthoryear{<author>}{<year>]{ref:?}
%    <ref. entry>
% \bibitem[\protect\citeauthoryear{<author>}{<year>]{ref:?}
%    <ref. entry>
% \end{thebibliography}

\end{document}